\documentclass[aps,prl,reprint,superscriptaddress, longbibliography]{revtex4-2}
\usepackage []{inputenc}
\usepackage{siunitx}
\usepackage{amsmath,bm}
\usepackage{amsbsy}
\usepackage{amssymb}
\usepackage{mathptmx, textcomp}
\usepackage{color}
\usepackage{xcolor}
\usepackage{braket}
\usepackage{graphicx}
\usepackage{textcomp}
\usepackage{notes2bib}
\usepackage{textcomp}
\usepackage{amsmath}
\definecolor{blouge}{rgb}{0.5, .1, .6}
\definecolor{bl}{rgb}{0, .1, .6}
\definecolor{turquoise}{rgb}{0.251, 0.878, 0.816}
\usepackage[colorlinks=true, citecolor = bl, linkcolor = bl, urlcolor=bl, pdfborder={0 0 0}]{hyperref}
\DeclareSIUnit\gauss{G}
\usepackage{placeins}
\let\Oldsection\section
\renewcommand{\section}{\FloatBarrier\Oldsection}

\usepackage[commentmarkup=uwave]{changes}

\definechangesauthor[color=blue]{dbo}

\begin{document}
\title{In-situ measurements of light diffusion in an optically dense atomic ensemble}

\author{Antoine Glicenstein}
\author{Apoorva Apoorva}
\author{Daniel Benedicto Orenes}
\author{Hector Letellier}
\author{\'{A}lvaro Mitchell Galv\~ao de Melo}
\author{Raphaël Saint-Jalm}
\author{Robin Kaiser}
\email[]{robin.kaiser@inphyni.cnrs.fr}
%\homepage[test]{Your web page}
%\thanks{}
\affiliation{Université Côte d’Azur, CNRS, Institut de Physique de Nice, 06200 Nice, France}

\begin{abstract}
This study introduces a novel method to investigate in-situ light transport within optically thick ensembles of cold atoms, exploiting the internal structure of alkaline-earth metals. A method for creating an optical excitation at the center of a large atomic cloud is demonstrated, and we observe its propagation through multiple scattering events. In conditions where the cloud size is significantly larger than the transport mean free path, a diffusive regime is identified. We measure key parameters including the diffusion coefficient, transport velocity, and transport time, finding a good agreement with diffusion models. We also demonstrate that the frequency of the photons launched inside the system can be controlled. This approach enables direct time- and space-resolved observation of light diffusion in atomic ensembles, offering a promising avenue for exploring new diffusion regimes.
\end{abstract}
\maketitle

Light propagation in disordered media is a complex problem that is encountered is various fields such as biomedical imaging \cite{Tempany2001,Davis2014,Weissleder2015}, solid-state physics \cite{fonda1992,gonis1999}, earth sciences \cite{Speziale2014,van2003,bluestein2022}, LIDAR-based sensing \cite{Wang2021}, or imaging through turbid media \cite{gu2015,sudarsanam2016,scheffold2012,lu2017}. Coherent wave propagation through such media features both attenuation and wavefront distortion arising from the optically dense and disordered character of the media \cite{carminati_principles_2021}. These two phenomena make the tasks of optically probing their characteristics or imaging through them more difficult, even if advanced techniques such as wavefront shaping \cite{Rotter2017,Yoon2020}, collection of ballistic or single-scattered photons \cite{kang_imaging_2015,zheng_ballistic_2021} or non-linear microscopy \cite{barad_nonlinear_1997,helmchen_deep_2005,evans_coherent_2008} allow to probe inside dense media in ranges up to 10-15 times the scattering mean free path.

However, disorder does not only hamper the observation of physical phenomena but, if well understood, it can be leveraged to bring in new and interesting physical effects such as optical transparency and superdiffusive light transport
\cite{Vynk2023}, coherent control of light waves \cite{Rotter2017}, or development of more efficient lasers \cite{Silva2024}. In particular, disorder can trigger a transition between the regime of diffusive transport of light and a regime where the waves are localized, also called Anderson localization \cite{Anderson1958}. This wave phenomenon has been experimentally observed for cold atoms and acoustic waves \cite{billy_direct_2008,chabe_experimental_2008,hu_localization_2008} but its observation is still lacking for light in three-dimensional materials \cite{van_der_beek_light_2012,sperling_can_2016,skipetrov_red_2016}.

Cold atomic experiments are a powerful platform to investigate light scattering in dense media \cite{Labeyrie1999,Pellegrino2014,Jennewein2016,Glicenstein2020,Wilkowski04,Kupriyanov2006}. First, the toolbox of atomic physics allows a reliable preparation of samples with well-controlled parameters and geometry. Second, the low temperatures achievable suppress the Doppler effect associated with atomic motion and provide samples with large optical densities with no source of dephasing, such as phonons in condensed matter systems. Third, the absence of inelastic processes during photon scattering simplifies the theoretical modelling of such samples and the analysis of experimental data.

In this letter, we use a laser-cooled 3D sample of Yb atoms as an optically dense medium with controllable mean free path and scattering cross-section to investigate light transport. In contrast with previous studies that excite their samples with light propagating through the boundaries of the medium, we introduce a new method to directly prepare and probe the deep core of our optically thick atomic sample. This technique maps the protocol of ultracold atoms experiments \cite{billy_direct_2008,jendrzejewski2012} onto photons. Indeed, whereas in the ultracold atom experiments the disorder can be switched on around the initial cloud (by adding the speckle induced random potential after the release of the initial Bose-Einstein condensate), in our case, the photons released in the center do not 'see' the atoms during their 'preparation' but become later resonant with the surrounding atoms. Using the toolbox of atomic physics, we first optically create a small excitation in the center of a large cloud of atoms and then follow its propagation in the cloud by monitoring the internal state of the atoms. This allows us to resolve both in time and in space the dynamics of the propagation of light in this medium and we circumvent the necessity to detect the scattered photons from outside of the cloud which requires to take into account the effect of their propagation through the whole cloud as in previous studies \cite{Labeyrie2003_prl,Labeyrie2005}. We report the measured diffusion coefficient as a function of the mean free path, and obtain an excellent agreement with theoretical predictions from a diffusion model.

The demonstration of this novel technique and its quantitative understanding opens the way to the study of new scattering regimes, for example where the light-induced dipole-dipole interaction may change the transport properties of the media \cite{Cherroret2016}. Our method may also constitute a progress towards the experimental observation of Anderson localization of light in 3D, where the observation of photons escaping the medium is precisely impossible. In particular, an initial excitation in the center of the disordered system avoids the problem of boundary effects, as the localized states are only well isolated from leakage when located several localisation lengths away from the boundary \cite{Tiggelen84,Payne2010}.

\begin{figure}
\includegraphics[width=\linewidth]{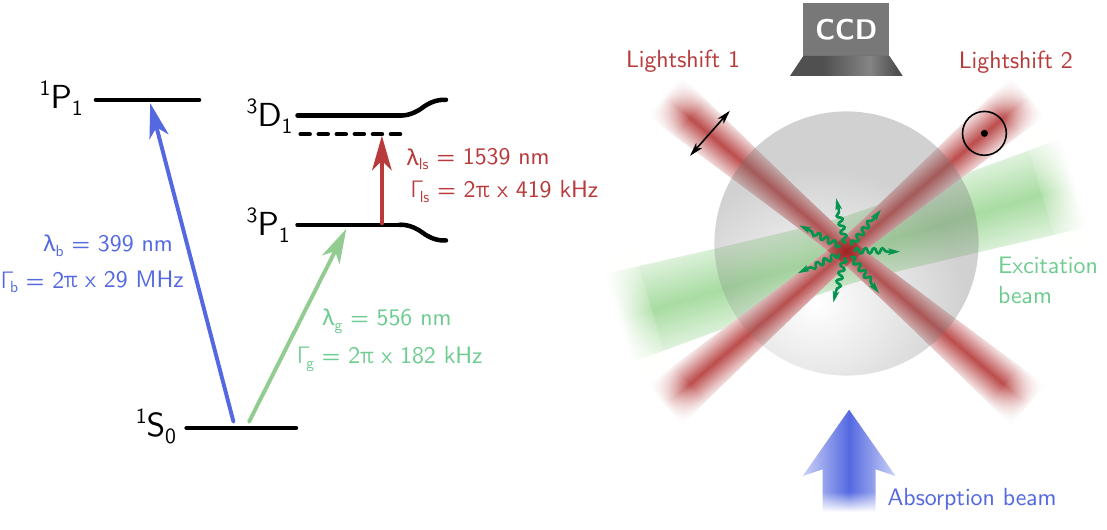} 
\caption{a)  Atomic transitions and levels used. b) Scheme of the experimental setup. A large dilute cloud of cold  $^{174}\mathrm{Yb}$ atoms is prepared in the center of a vacuum cell. Excitations in the "narrow" $^3P_{1}$ state (green, $\lambda_{\mathrm{g}}$) are observed in space and time using absorption imaging on the "broad" transition (blue,    $\lambda_{\mathrm{b}}$). The excitation beam is roughly perpendicular to the absorption imaging beam. The cloud is spatially dressed by two laser beams near-resonant with the ${}^{3}P_{1}-{}^{3}D_{1}$ transition (wavelength $\lambda_{\mathrm{ls}}$).}
\label{Fig1}
\end{figure} 

%\paragraph{Description of the experimental system:}
A detailed description of the experimental setup can be found in previous works \cite{letellier_piegeage_2024,Letellier2023,deMelo2024,deMelo_PhD_2024}.
Using a single slowing beam and a magneto-optical trap (MOT) on the dipole allowed transition ${}^{1}S_0 - {}^{1}P_1$ with wavelength $\lambda_{\mathrm{b}}=\SI{399}{\nano\meter}$ and linewidth $\Gamma_{\mathrm{b}}=2\pi\times\SI{29}{\mega\hertz}$ (called \emph{broad} transition in the following), atoms are efficiently cooled and trapped from an atomic beam of $^{174}\mathrm{Yb}$. These atoms are then transferred to a second MOT using the $\emph{narrow}$ intercombination transition ${}^{1}S_0 - {}^{3}P_1$ with wavelength 
$\lambda_{\mathrm{g}}=\SI{556}{\nano\meter}$ and linewidth $\Gamma_\mathrm{g} = 2\pi\times\SI{182}{\kilo\hertz}$ (see Figure \ref{Fig1}a). This procedure allows to create atomic clouds containing up to $ 3\times10^8$ atoms at a typical temperature $T\sim\SI{15}{\micro\kelvin}$, and with a repetition rate of $\SI{1}{\hertz}$.

To study the transport properties of the light in the medium, it is necessary to control the transport mean-free path and the on-resonance optical depth $b_0$ to go from the dilute regime where $b_0 \ll 1$ to the multiple scattering regime requiring $b_0 \gg 1$. To achieve that, we choose the initial atom number $N$ and allow the cloud to freely expand during a time-of-flight (tof) of up to $\SI{25}{\milli\second}$ \footnote{We have verified  that the transport properties are governed by the mean-free-path and not by the number of emitters: two clouds with the same optical depth but a factor two difference in atom number give the same results as long as the mean-free-path is much smaller than the cloud size.}. This varies the radius $r_0$ of the cloud, and in turn the on-resonance optical depth $b_0=(3 N) / (4\pi^2) (\lambda/r_0)^2$ and the mean-free path via $l=\sqrt{2\pi}r_0/b_{0}$. We characterize  our atomic cloud using absorption imaging on the $\emph{broad}$ transition, as represented on Figure \ref{Fig1}b). We convert this optical depth to the corresponding optical depth for the \emph{narrow} transition by $b_{0,\mathrm{g}} = b_{0,\mathrm{b}} \left( \lambda_{\mathrm{g}}/\lambda_{\mathrm{b}} \right)^2$ (with an additional minor correction due to Doppler broadening \cite{Hu2022,supp}). This allows us to explore atomic clouds with peak optical depths on the narrow transition ranging from $1$ to $30$.

\begin{figure} 
\includegraphics[width=\linewidth]{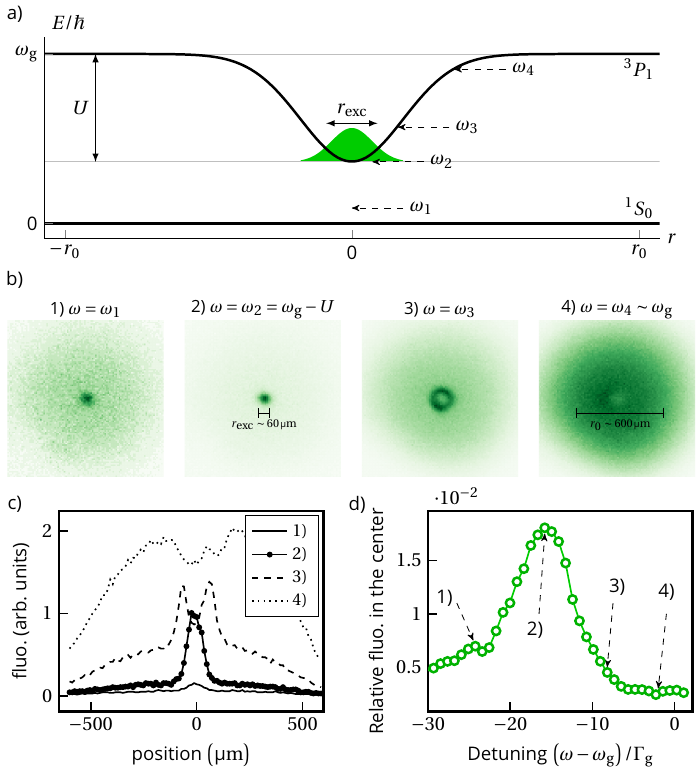} 
\caption{ a) Energy of  the dressed ${}^{3}P_{1}$ state. A lightshift $U \gg \Gamma_{\mathrm{g}}$ is applied in the center of the MOT (of size $r_0$) together with an excitation beam of frequency  $\omega$. When $\omega= \omega_2=\omega_{\mathrm{g}}-|U|/\hbar$, only atoms in volume with a size $r_{\mathrm{exc}}$ are excited. b) Single-shot fluorescence images (normalized, false color) illustrating the  excited regions as a function of the excitation beam frequency: $\omega_1\sim \omega_{\mathrm{g}} - |U|/\hbar - 10\Gamma_{\mathrm{g}}$, $\omega_2 = \omega_{\mathrm{g}} - |U|/\hbar$, $\omega_3 \sim \omega_{\mathrm{g}} - |U|/\hbar + 6\Gamma_{\mathrm{g}}$ and $\omega_4\sim \omega_{\mathrm{g}}$ (see panel d)).  The cloud used has an optical depth of $b_{0,\mathrm{g}} \simeq 20 $. c) Vertical cuts along the images of panel b). d) Relative fluorescence in a square of dimensions $\sim r_{\mathrm{exc}}$ in the center compared to the total fluorescence. This relative fluorescence displays a maximum when $ \omega = \omega_2 = \omega_{\mathrm{g}} - |U(r=0)|/\hbar$ (case 2).}
\label{Fig2}
\end{figure}

%\paragraph{Creating and detecting a local excitation:}
To prepare a local excitation in the core of our sample, even when optically dense, we use the toolbox of atomic physics. As depicted in Figure \ref{Fig1}a, a position-dependent light shift is applied using two laser beams far-off resonance with the ${}^{3}P_{1}-{}^{3}D_{1}$ transition ($\lambda_{\mathrm{ls}}=\SI{1539}{\nano\meter}$, Figure \ref{Fig1}a). The two beams are focused to a waist $w_{\mathrm{ls}}$ $= \SI{300}{\micro\meter}$ crossing at the center of the MOT with an angle $ \sim \SI{90}{\degree}$ and with orthogonal polarizations. The induced light shift of the narrow transition scales as $U(r) \propto \Gamma_{\mathrm{ls}} I_{\mathrm{ls}} \left( r \right)/{\Delta_{\mathrm{ls}}} $ \cite{grimm_optical_2000} where $r$ is the distance from the center of the MOT, $\Gamma_{\mathrm{ls}}$, $\Delta_{\mathrm{ls}}$ and $I_{\mathrm{ls}}$ are respectively the resonant scattering rate, detuning and laser intensity of the dressing beam. Atoms are initially prepared in the ground state ${}^{1}S_{0}$. They are then optically pumped using a single beam (called \emph{excitation beam} in the following) with a detuning $ \omega - \left(\omega_\mathrm{g} -|U(r=0)|/\hbar \right)$ with respect to the dressed  transition, where $\omega_{\mathrm{g}}=2\pi c/\lambda_{\mathrm{g}}$. We ensured that the waist of the beam at the position of the MOT is large enough such that it can be considered as a plane wave when interacting with the atoms in the sample. When $\omega=\omega_2 = \omega_\mathrm{g} -|U(r=0)|/\hbar $, this light is resonant with the atoms in the center of the cloud, as depicted Figure \ref{Fig2}a. This beam excites atoms located within a radius $r_\mathrm{exc}\sim w_{\mathrm{ls}}\sqrt{\hbar\Gamma_{\mathrm{g}}/(2|U|)}$ that can be made much smaller than the cloud radius $r_0$ by increasing $|U|$ or reducing $w_{\mathrm{ls}}$. We choose the beam waist such that $r_0/r_\mathrm{exc} \sim 10$, securing an initial excitation much smaller than the whole cloud, but with sufficient signal to be collected. Knowing that $\Gamma_\mathrm{ls} = 2\pi\times\SI{419}{\kilo\hertz}$\cite{Beloy2012}, we tune the dressing laser frequency to about $ \Delta_{\mathrm{ls}} \sim \SI{-6}{\giga\hertz}$ from the ${}^{3}P_{1}-{}^{3}D_{1}$ transition to ensure $\Delta_{\mathrm{ls}}\gg \Gamma_{\mathrm{ls}}$ and a negligible scattering. The resulting lightshift in the center is $|U(r=0)| \sim 15 \hbar\Gamma_{\mathrm{g}}$.

 In order to illustrate this technique in a qualitative way, we first measure  $U(r)$ by imaging the fluorescence emitted by the cloud at $\SI{556}{\nano\meter}$ after a pulse of the excitation beam of duration $t_{\mathrm{fluo}} = \SI{1}{\milli\second}$, as a function of the excitation frequency $\omega$. Typical images are shown in Figure \ref{Fig2}b for various frequencies of the excitation beam with respect to the dressed transition, and for a cloud of optical depth $b_{0,\mathrm{g}} \simeq 20$. Figure \ref{Fig2}c shows a vertical cut through those images. When scanning $\omega$, the region of space where the atoms are excited varies. Far from resonance (case 1), atoms are weakly excited with an almost uniform probability. At $\omega= \omega_{\mathrm{g}} - |U|/\hbar$ (case 2), only atoms in a volume of radius $r_{\mathrm{exc}} \sim \SI{60}{\micro\meter}$ are excited. Close to the bare resonance $\omega=\omega_{\mathrm{g}}$, almost all the atoms of the cloud are excited, except in the center (case 4). In the intermediate frequency range, a spherical shell around the center of the cloud is excited (case 3). $U(r=0)$ is measured considering the relative fluorescence in the center compared to the total fluorescence, see Figure \ref{Fig2}d. Indeed, even though the total fluorescence increases near  the non-dressed resonance, the relative fluorescence is maximum at $\omega=\omega_2$. This protocol demonstrates that we are able to excite atoms in the center of an optically dense medium, and gives a rough estimate of the excitation size. 
 
 However, since the detection time $t_{\mathrm{fluo}}$ is much larger than the evolution time $ \tau_{\mathrm{g}} = 1/\Gamma_{\mathrm{g}} = \SI{866}{\nano\second}$, this method does not give access to high resolution temporal dynamics. We therefore use another method based on the broad transition to image the atoms in the ground state. It can be sufficiently short so that the dynamics of the photons at $\lambda_\mathrm{g}$ is frozen. The excitation pulse is applied for $\SI{10}{\micro\second}$ to ensure that a steady-state is reached, with an intensity $I \sim I_{\mathrm{sat}}$ \footnote{We observe that results do not change significantly with the power of the excitation beam. We use an intensity of about $I_{\mathrm{sat}}$ to maximize the signal to noise ratio while having observed no important change in the results presented here.}. For the absorption imaging, pulses of \SI{150}{\nano\second} are then applied on the broad transition, temporally shaped by acousto-optic modulators. This time range is sufficient to resolve the dynamics of the narrow transition while using standard electronics to control the pulses. Absorption images are recorded using a high speed CMOS camera \footnote{Camera model: Andor Zyla 5.5}.

\begin{figure} 
\includegraphics[width=\linewidth]{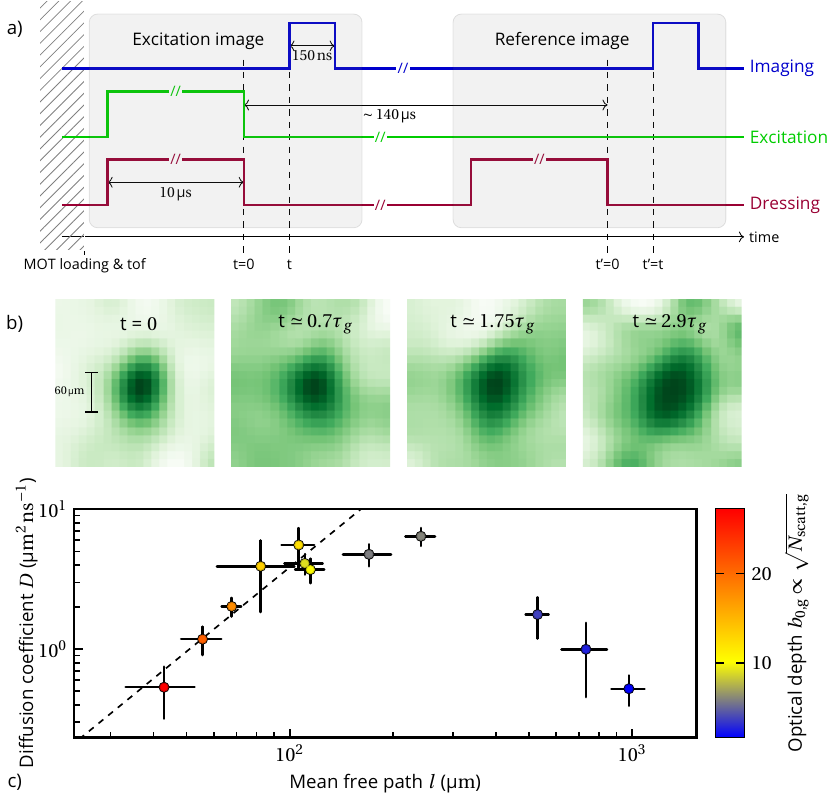} 
\caption{a) Chronogram of the protocol. After MOT loading and free expansion, two absorption images are taken with and without the excitation beam. A "dark" image without light is subtracted on both. An experimental image is the ratio of the two absorption images. b) Example of the diffusion of light: snapshots for 4 different times between 0 and $\sim 3\tau_{\mathrm{g}}$ showing the spreading of the excitation. All images are normalized by their maximum and a gaussian smoothing  (size 1 pixel) is applied for display. Details of how images are produced are given in the text. c) Diffusion coefficient $D$ as a function of the mean-free path $l$. The colormap shows the optical depth for all ensembles prepared, proportional to the square root of the number of scattering events. The dashed line represents the theoretical expression $D = l^2/(3\tau_{\mathrm{g}})$. } \label{Fig3}
\end{figure} 

In order to follow the time and space dynamics of the excitation, we compare absorption images with (labelled "\emph{exc}") and without the presence of the excitation (labelled "\emph{ref}") as a function of time, characterized by the recorded imaging light intensities $I_{\mathrm{exc/ref}}(x,y,t) =I_0 \exp{[-b_{\mathrm{exc/ref}}(x,y,t)]}$, where $b_{\mathrm{exc/ref}}(x,y,t)$ is the optical depth for the broad transition along $z$ at position $(x,y)$. This optical depth can be expressed as
\begin{equation}
b_{\mathrm{exc/ref}}(x,y,t)=\int{ \sigma_{\mathrm{sc,b}} \rho_{\mathrm{exc/ref}}(x,y,z,t)\ \mathrm{d}z},
\end{equation}
with $\rho_{\mathrm{exc/ref}}$  the atomic density in the state ${}^1S_0$ and $\sigma_{\mathrm{sc,b}}$ the scattering cross-section for the broad transition.

When comparing the two images with and without the excitation, we measure a smaller optical density in the former, since some atoms are in the ${}^3P_1$ state that is insensitive to the absorption imaging beam. We define $\rho_\mathrm{e}$ the density of these excited atoms by
\begin{equation}\label{eq:density_e}
\rho_\mathrm{e} = \rho_\mathrm{ref} - \rho_\mathrm{exc}.
\end{equation}
To measure this density we compute the ratio $I_{exc}/I_{ref}$, and if $b_\mathrm{ref} - b_\mathrm{exc} \ll 1$, we find that
\begin{equation}
\frac{I_{\mathrm{exc}}(x,y,t)}{I_{\mathrm{ref}}(x,y,t)} \approx 1 + \sigma_{\mathrm{sc,b}} \int{\rho_\mathrm{e} (x,y,z,t) \ \mathrm{d}z}.
\end{equation}

We can thus measure the space-time dynamics of the density of the excited atoms integrated along the direction of imaging by computing the ratio of two recorded absorption images on the broad transition. If we assume a gaussian density for the excited atoms, $\rho_\mathrm{e}(x,y,z,t) = \rho_{\mathrm{e},\perp}(x,y,t)\rho_{\mathrm{e},z}(z,t)$, therefore:
\begin{equation}
\frac{I_{\mathrm{exc}}(x,y,t)}{I_{\mathrm{ref}}(x,y,t)} \approx 1 + \sigma_\mathrm{sc,b} K(t) \rho_{\mathrm{e},\perp}(x,y,t),
\end{equation}
where $K(t) = \int{\rho_{\mathrm{e},z}(z,t)\ \mathrm{d}z}$. The spatial resolution is given by the resolution of the absorption images, which is $\SI{22}{\micro\meter}$. For equation \ref{eq:density_e} to hold experimentally, both absorption images are taken from the same cloud in order to reduce the effect of random loading and residual atomic temperature.  The time between the images is reduced to $\SI{140}{\micro\second}$. The imaging sequence is described in Figure \ref{Fig3}a. We then subtract the remaining background due to the free fall of atoms between the two images, calibrated for each set of parameters. Each image is averaged over at least 50 absorption images. See supplementary materials for detailed explanation \cite{supp}.

We typically record data up to $3\tau_{\mathrm{g}}$ after the end of the excitation pulse, and monitor the evolution of the density of excited atoms, which provides a proxy for the dynamics of the resonant photons in the optically dense cloud. Examples of data are shown Figure \ref{Fig3}b, for different times. A 2D-gaussian fit is used to extract an amplitude $A(t)$ and  width $r(t)$ characterizing the density of excited atoms. The diffusion coefficient of the photons is computed from the slope of a linear fit of $r^2 (t)$ \cite{supp}. Results are shown Figure \ref{Fig3}c, as function of the transport mean-free path $l=\sqrt{2\pi}r_0/b_{0,\mathrm{g}}$, where $b_{0,\mathrm{g}}$ is the optical depth of the cloud for the photons at $\lambda_\mathrm{g}$. The colormap shows $b_{0,\mathrm{g}}$ at the center of the cloud used for each point. It is related to the average number of scattering events experienced by the excitation $N_{\mathrm{scatt,g}}$ by $b_{0,\mathrm{g}}\propto \sqrt{N_{\mathrm{scatt,g}}}$. In the limit $N_{\mathrm{scatt,g}} \gg 1$ (multiple scattering regime), a diffusive behavior is expected. We observe that for a mean-free path smaller than about $\SI{100}{\micro\meter}$, corresponding  to   $b_{0,\mathrm{g}} \geq 8$, the measured diffusion coefficient $D$ follows the expected law $D = l^2/(3\tau_{\mathrm{g}})$, shown with a dashed line Figure \ref{Fig3}c. For lower $b_{0,\mathrm{g}}$, it deviates from the diffusive regime as the excitation leaves the cloud before being sufficiently scattered, with a "ballistic" regime for vanishing $b_{0,\mathrm{g}}$.
\begin{figure} 
\includegraphics[width=\linewidth]{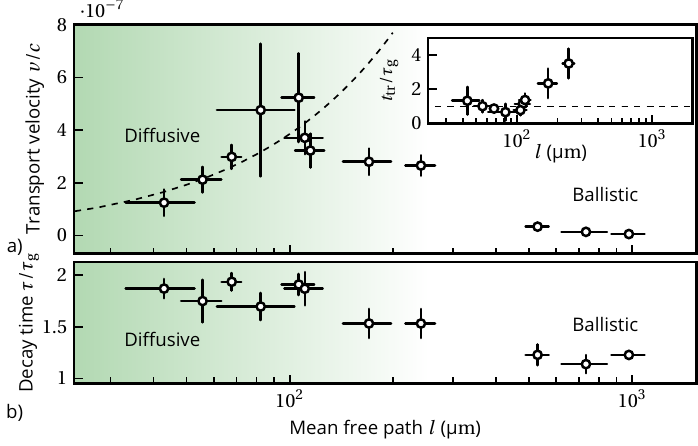} 
\caption{a) Transport velocity $v/c$ as a function of the mean-free path $l$. The dashed line represents the theoretical expression $v = \frac{l}{\tau_{\mathrm{g}}}$. Inset : transport time as a function of $l$. b) Excitation decay time $\tau/\tau_{\mathrm{g}}$ as a function of the mean-free path $l$.} \label{Fig4}
\end{figure} 
From the same data, we extract the transport velocity $v=3 D/l$ and the transport time $t_{\mathrm{tr}} = l^2/(3 D)$. Results are shown Figure \ref{Fig4}a. We note that the transport velocity is almost seven orders of magnitude smaller than the speed of light in vacuum, and two orders of magnitude slower than results obtained previously with alkali atoms due to the longer lifetime of the narrow transition \cite{Labeyrie2003_prl}. We have verified that $t_{\mathrm{tr}}=\tau_{\mathrm{g}}$, as shown in the inset of Figure \ref{Fig4}a \footnote{As transport time $t_{\mathrm{tr}}$ is computed from the diffusion coefficient, its value differs from $\tau_{\mathrm{g}}$ outside from the diffusive regime but has no physical meaning. This remark also holds for the transport velocity.}. $v$ can be expressed as  $v = l\:\Gamma_{\mathrm{g}}$, which shows the advantage of using a narrow transition to explore transport properties in optically dense media. Finally, we define the excitation decay time $\tau$ as the characteristic time of the decay of the amplitude $A(t)$. This time is obtained using an exponential fit of $A(t)$, which turned out to be a good approximation for each dataset. In the purely ballistic regime, this time should be equal to $\tau_{\mathrm{g}}$, which is shown Figure \ref{Fig4}b. In the diffusive regime, $\tau$ increases to $\approx 2$. However, these obtained decay times are much smaller than the decay times measured  with the transmitted diffuse light in a radiation trapping experiment \cite{Labeyrie2003_prl,Labeyrie2008}, even taking the finite atomic temperature into account \cite{Labeyrie2005}. We attribute this discrepancy to the fact that here we measure only the early decay, thus not the lowest Holstein mode, which scales like $b_{\mathrm{g}}^2$ \cite{Labeyrie2003_prl}. This decay is not accessible with our method because the experimental signal to noise becomes too small after a few $\tau_{\mathrm{g}}$. 

In addition to the above discussed control of the spatial extent of the initial excitation, this technique also allows to control the frequency of the photons launched into the disordered cloud of cold atoms. Indeed, for the results presented before we always turned off the dressing laser after the preparation of the initial excitation. In this situation, the atoms return to their bare eigenstates and photons are emitted  at $\omega_{\mathrm{0,g}}$, which is validated experimentally by the fact that the observed diffusion matches perfectly with the mean-free path measured on-resonance. However, if the dressing laser is kept on, the frequency of the photons emitted from the initially prepared excitation at the core of the sample will be shifted accordingly. In order to illustrate this effect, we have performed an experiment where we kept the dressing laser at its initial value when switching off the narrow excitation laser. We thus launch photons at $\omega=\omega_2 = \omega_{\mathrm{0,g}} - 15 \Gamma_{\mathrm{g}}$. These photons being very far detuned from resonance for the atoms of the rest of the cloud, they are in the ballistic regime. In this case, $D$ goes to zero.

In this letter, we have introduced a new method for the in-situ study of light transport in optically thick ensembles of atoms, based on the toolbox of atomic physics and alkaline-earth atoms. A protocol to create an on-resonance excitation in the center of the cloud is demonstrated, and we characterize the space-time evolution of this excitation. A diffusive regime has been identified, due to multiple scattering. The obtained diffusion coefficients as a function of the mean-free path are in agreement with the expected values, as are the transport time and transport velocity. These results benchmark the method, opening the road to the study of other scattering regimes. In particular, this study has been performed in a dilute ensembles ($\rho \sim \SI[parse-numbers=false]{10^{10-12}}{\per\cubic\centi\meter}$) but can be extended to the dense regime, where light-induced dipole-dipole interaction may change the diffusive behavior \cite{Cherroret2016}. It phase control of the photons can be achieved, this will also open the road to studies including directed in-situ excitation as for instance required for Coherent Forward Scattering \cite{Karpiuk2012}. The method is a promising progress towards the experimental study of Anderson localization of light in atomic gas: a finite size system can only have its core in the localized regime, and the regions close to the boundaries stay diffusive, which justifies the advantage of tools that directly probe the core of a sample.

\begin{acknowledgments}
This work was performed in the framework of the European project ANDLICA, ERC Advanced grant No. 832219. D.B.O is supported by European Union’s Horizon 2020
research and innovation program under the Marie Skłodowska-Curie grant
agreement No. 10110529.

\end{acknowledgments}


%\bibliography{Biblio}

\end{document}